\documentclass{iaus}
\usepackage{graphicx}

\title[Parallel electric field generation] 
{A minimal model of parallel electric field generation in a transversely inhomogeneous plasma}

\author[David Tsiklauri]   
{David Tsiklauri}

\affiliation{Institute for Materials Research,
University of Salford, Greater Manchester, M5 4WT, United Kingdom.}

\pubyear{2008}
\volume{247}  
\pagerange{xxx--xxx}
\date{?? and in revised form ??}
 \jname{Waves \& Oscillations in The Solar
Atmosphere: \\Heating and Magneto-Seismology} \editors{R.
Erd\'{e}lyi \& C.~A. Mendoza-Brice$\tilde{n}$o,
eds.}
\begin{document}

\maketitle

\begin{abstract}
The generation of parallel electric fields by the propagation of ion cyclotron waves
 (with frequency 0.3 $\omega_{ci}$) in the plasma
with a transverse density inhomogeneity was studied. Using two-fluid, cold plasma linearised equations, it was shown 
for the first time that, in this particular context,  $E_{\parallel}$ generation 
can be understood by an analytic equation that couples $E_{\parallel}$  to the transverse electric 
field of the driving ion cyclotron wave.
It was proven that the minimal model required to reproduce the previous kinetic simulation results
of $E_{\parallel}$ generation [Tsiklauri et al 2005, G\'enot et al 2004] is the 
two-fluid, cold plasma approximation in the linear regime. 
By considering the  numerical solutions it was also shown that the cause 
of $E_{\parallel}$ generation is  the electron and ion flow separation 
induced by the transverse density inhomogeneity.
We also investigate how $E_{\parallel}$ generation is affected by the mass ratio and found
that amplitude attained by $E_{\parallel}$ decreases linearly as inverse of the mass ratio $m_i/m_e$. 
For realistic mass ratio of $m_i/m_e=1836$, such 
empirical scaling law, within a time corresponding to 3 periods of the driving ion cyclotron wave, 
is producing  $E_{\parallel}=14$ Vm$^{-1}$ for solar coronal
parameters. Increase in mass ratio does not have any effect on final 
parallel (magnetic field aligned) speed attained by electrons. However, parallel ion
velocity decreases linearly with inverse of the mass ratio $m_i/m_e$. 
These results can be interpreted as following: (i) ion dynamics plays no role in the $E_{\parallel}$
generation; (ii) $E_{\parallel} \propto 1/m_i$ scaling
is caused by the fact that $\omega_d = 0.3 \omega_{ci} \propto 1/m_i$ is decreasing 
with the increase of ion mass, 
and hence the electron fluid can effectively "short-circuit" (recombine with) the slowly 
oscillating ions, hence producing smaller $E_{\parallel}$.
\keywords{waves, hydrodynamics, Sun: atmosphere, Sun: Corona, Sun: oscillations}
\end{abstract}

The generation of parallel electric fields in inhomogeneous plasmas is a generic
topic, which is of interest in a variety of plasma phenomena such as 
particle acceleration in Solar and stellar flares \cite{lf05}, auroral acceleration
region and current sheets in the Earth magnetosphere (see refs. in \cite{sl06}), laboratory plasma reconnection 
experiments \cite{yam97} and many more.
In situ and remote observations of accelerated particles often show
parallel electric fields in localised double layers, charge holes or U-shaped voltage drops.

The issue of $E_{\parallel}$ generation by the propagation of ion cyclotron waves
in the plasma with a transverse density inhomogeneity is
discussed in more detail in \cite{t07}. 

We start from two-fluid, cold (ignoring thermal
pressure) plasma linearised equations \cite{kt73}:
\begin{equation}
\partial_t \vec V_e=-(e/m_e)\left(\vec E + \vec V_e \times \vec B_0 /c \right),\\
\end{equation}
\begin{equation}
\partial_t \vec V_i=+(e/m_i)\left(\vec E + \vec V_i \times \vec B_0 /c \right),
\end{equation}
\begin{equation}
\partial_t \vec B= -c \nabla \times \vec E,
\end{equation}
\begin{equation}
\partial_t \vec E = c \nabla \times \vec B - 4 \pi n e (\vec V_i - \vec V_e).
\end{equation}
Hereafter subscripts under $\partial$ denote partial derivative with respect to that
subscript. Uniform, background magnetic field, $B_0$ is in $z$-direction.
Density profile is specified as a ramp, 
$n(x)=n_0\left(1+3 \exp \left[-[(x-100\delta)/ (20 \delta)]^6 \right]\right)$
in which the central region (along $x$-direction, i.e. across $z$),
is smoothly enhanced by a factor of 4, and there
are the strongest density gradients having a width of about $20 \delta$
around the points $x = 81 \delta$ and $x = 119\delta$.
Here $\delta=c/\omega_{pe}$ is the (electron) skin depth, which is a
unit of  grid in our numerical simulation.
We use 2.5D description meaning that we keep all three, $x,y,z$ components
of all vectors, however spatial derivatives $\partial / \partial y \equiv 0$.

In order to derive  the equation that describes  $E_{\parallel}=E_z$
generation, we write Eqs.(1)-(4) in $x,y,z$ component form. Omitting details of the
calculation we present the final result:
\begin{equation}
\left(\partial^2_{tt} - c^2 \partial_{xx}^2 + \omega_{pi}^2+ \omega_{pe}^2 \right) E_{\parallel}=
-c^2 \partial^2_{zx} E_x.
\end{equation}
Also, a similar calculation enables us to obtain the equation describing the dynamics 
of driving transverse electric field $E_x$ on an ion cyclotron wave:
$$
\left(\partial^2_{tt} - c^2 \partial_{zz}^2 + \omega_{pi}^2+ \omega_{pe}^2 \right) E_x=
$$
\begin{equation}
-c^2 \partial^2_{zx} E_{\parallel}-
\omega_{pi}^2(m_i/e)\omega_{ci}V_{iy}-\omega_{pe}^2(m_e/e)\omega_{ce}V_{ey}.
\end{equation}
Note that Eq.(6) also describes the feedback of the generated $E_{\parallel}$ on the
driving transverse electric field $E_x$ (see the first term on the right-hand-side).
Here the notation is standard: $\omega_{pe}=\sqrt{4 \pi n_0 e^2/m_e}$ and $\omega_{pi}=\sqrt{4 \pi n_0 e^2/m_i}$  are
electron and ion plasma frequencies;  $\omega_{c(e,i)}=eB_0/(m_{(e,i)}c)$ are respective cyclotron frequencies.

 In order to solve Eqs.(1)-(4) numerically we use the following normalisation:
$t=\tilde t \omega_{pe}^{-1}$, $V_{x,y,z}=\tilde V_{x,y,z} c$, $E_{x,y,z}=\tilde E_{x,y,z} (m_e c \omega_{pe} / e)=
\tilde E_{x,y,z} E_0$,
$B_{x,y,z}=\tilde B_{x,y,z} B_0$, and $(x,y,z)= c/\omega_{pe} (\tilde x, \tilde y, \tilde z)$.
In what follows we omit tilde on the dimensionless quantities.
The $(x,z)$ simulation 2D box size is $200 \delta \times 2500 \delta$. Since we fix background 
plasma number density at $10^9$ cm$^{-3}$ (typical value for the solar corona), 
$\omega_{pe}$ is then $1.784 \times 10^9$ rad s$^{-1}$ and the simulation box size is 
$33.6$ m in $x$-  and $420.5$ m in $z$-direction. $B_0$ was fixed at 101.5 Gauss 
(typical value for the solar corona), which gives $\omega_{ce} / \omega_{pe}= 1$.
$m_i / m_e$ ratio was varied as: 45.9, 91.8, 183.6 to 262.286 (realistic one is 1836).
These values correspond to $1/40,\; 1/20, \; 1/10$ and $1/7$-th of the realistic value
respectively.
This yields respectively: 
$\omega_{ci} / \omega_{pi} = B_0/(c \sqrt{4 \pi n_i m_i})=V_A / c=1/\sqrt{m_i/m_e}=0.148, \;0.104,\;0.074$
and $0.062$ for $x\leq 70$
and  $x\geq 130$ (realistic $\omega_{ci} / \omega_{pi} =V_A / c$ is 0.023).
Here parameters are similar to e.g. \cite{tss05a,tss05b}, except for
far more realistic mass ratios. Note that the simulation parameters are still
somewhat artificial. 
Full kinetic, Particle-In-Cell (PIC) simulations employed in \cite{tss05a,tss05b} or
in gyro-kinetic approach which uses guiding centre approximation  for electrons, whilst retaining ion particle-like dynamics \cite{glq99,glm04}
are computationally challenging. 
Thus, in those studies rather modest mass ratios e.g. 16 were used.
Note also, that since here we do not need to resolve electron 
thermal motions as we are only studying electromagnetic part of the 
problem ($E_{\parallel}$ generation) our unit of spatial grid size is $\delta=c/\omega_{pe}$, the (electron) skin depth.
While in full kinetic, PIC simulation \cite{tss05a,tss05b} the unit of grid has to be $\Delta=v_{th,e}/\omega_{pe}$.
Since in a PIC simulation typically $v_{th,e} / c =0.1$, in the present, two-fluid approach an equivalent to PIC numerical simulation 
requires $(\delta/ \Delta)^2=(c/v_{th,e})^2=10^2$ less grid points, thus it can be 100 times faster. 
For comparison a single run for mass ration 16 
in \cite{tss05a,tss05b} takes about 8 days on parallel, 32 dual-core 2.4 GHz Xeon processors,
similar run with mass ratio of 262 would have taken 4 months.
The numerical run presented here for the mass ratio of 262 takes 4 days with only one processor.

We solve {\it relativistic} version of
Eqs.(1)-(4) numerically with a specially developed and tested FORTRAN 90 code
which uses 4-th order centred spatial derivatives and 4-th order Runge-Kutta time marching. 
Although Alfv\'en speeds considered are at most $\approx 15$ \% of the speed of light for $m_i/m_e=45.9$,
relativistic effects were included. The simplest option becomes available in the linear
regime. In   \cite{kt73}, appendix I, paragraph 5, it was shown that the
relativistic equation of motion of a particle with charge $q$ and the rest mass $m_0$ 
can be written as
\begin{equation}
\frac{d}{dt}\vec V =\frac{q}{\gamma m_0} \left[\vec E +\frac{\vec V \times \vec B}{c} - 
\frac{\vec V (\vec V \cdot \vec E)}{c^2} \right],
\end{equation}
where $\gamma=(1-V^2/c^2)^{-1/2}$. As can be seen from the latter equation, in the linear regime, it
coincides with either Eq.(1) or (2) after substituting $m_{e,i} \to \gamma_{e,i} m_{e,i}$, where
$\gamma_{e,i}=(1-V_{e,i}^2/c^2)^{-1/2}$. Naturally, such simplified approach is only valid 
when there are no flows in the unperturbed state $V_0=0$. As can be seen below, largest attained velocities
in the simulation are those of electrons, and these do not exceed 3 \% of speed of light. Thus,
relativistic corrections play only a minor role. It should be noted, however we still retain the displacement current
in Eq.(4).
Note, also that the gradients in the code are resolved numerically to an appropriate precision (20 grid points across each gradient.)

\begin{figure}
\centering
\resizebox{6.5cm}{!}{\includegraphics{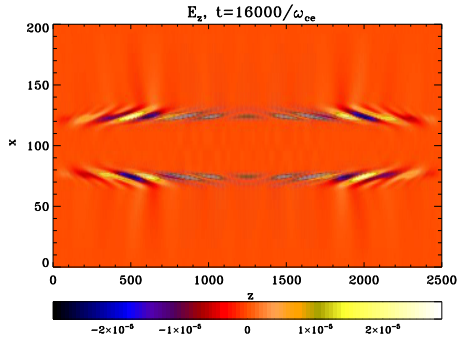} }
\caption[]{Contour plot of $E_z=E_{\parallel}$ at time  $t=16000 / \omega_{ce}$. Here $m_i/m_e=262$.}
\end{figure}

Initially all perturbations are set to zero, and we start driving the $z=1$ cell with the transverse magnetic fields of the form
$B_y  =  -0.05 \sin(\omega_d t) \left(1.0-\exp[-(t/(3.125 \omega^{-1}_{ci}))^2]  \right)$
and $B_x =  -0.05 \cos(\omega_d t) \left(1.0-\exp[-(t/(3.125 \omega^{-1}_{ci}))^2]  \right)$. 
As in \cite{tss05a,tss05b}, we fixed $\omega_d$ at 0.3 $\omega_{ci}$ (to avoid ion-cyclotron damping playing any role).
$\left(1.0-\exp[-(t/(3.125 \omega^{-1}_{ci}))^2]  \right)$ factor ensures that these driving $B_\perp$ fields ramp up to their maximal values
in time $t=3.125 \omega_{ci}^{-1}$.
Such driving with $B_\perp$ of 5\% of the background $B_0$ 
excites circularly polarised ion-cyclotron (IC) waves, 
these waves are often misquoted as Alfv\'en waves \cite{glq99,glm04,tss05a,tss05b}.
Although, in the frequency range $\omega \ll \omega_{ci}$ both left and right polarised 
IC waves tend to an Alfv\'en wave branch in the dispersion relation \cite{kt73},
at frequencies $\omega \simeq 0.3 \omega_{ci}$ the correct term ion-cyclotron wave instead should be used.
 In the considered problem $E_x$ and $B_y$ are both components of Alfv\'en (ion cyclotron) wave, so these can be used
interchangeably. Conventionally, Alfv\'enic and IC waves are more associated with magnetic field oscillation.
It is in kinetic, Particle-In-Cell simulations driving by electric field is 
required because particles respond to electric, rather than magnetic fields. In the 
two-fluid simulation this is not a requirement.

The generated at the left edge ($z=1$) IC waves propagate both in the directions of positive and negative $z$'s.
However, because of the periodic boundary conditions used (applied on all physical quantities) IC wave 
that travels to the direction of negative $z$'s (to the left)
re-appears on the right edge of the simulation box. 
The driving $B_\perp$ dynamics is not presented here (see for details \cite{t07}), but it resembles closely to the previous kinetic 
simulation results  \cite{glq99,glm04,tss05a,tss05b}.
As in all previous phase-mixing simulations Alfv\'en velocity is a function
of the transverse (to the background magnetic field) coordinate, $x$, i.e. $V_A=V_A(x)\propto 1/\sqrt{n(x)}$. Thus the 
IC wave middle portion travels slower than the parts close
to the simulation box edge. This creates progressively strong transverse gradients and hence smaller spatial scales.
If resistive effects are included (these are absent here), such a configuration usually produces greatly 
enhanced dissipation and IC wave amplitude decays in space as 
$\propto \exp(-z^3)$ \cite{tss05a,tss05b}. 

The $E_{\parallel}=E_z$ field snapshot at $t=16000 / \omega_{ce}$ is shown in Fig.(1). 

\begin{figure}
\centering
\resizebox{6.5cm}{!}{\includegraphics{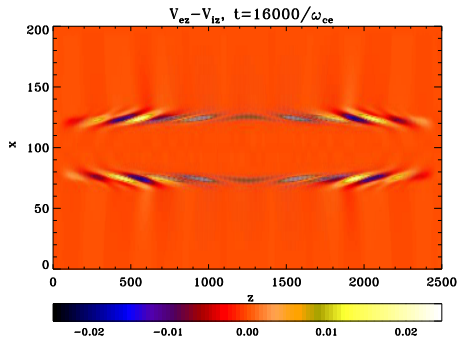} }
\caption[]{Contour plot of $(V_{ez}-V_{iz}) \propto j_z$ at time $t=16000 / \omega_{ce}$. Here $m_i/m_e=262$.}
\end{figure}

$E_{\parallel}$
is generated only in the regions of density gradients i.e. around $x = 81$ and $x = 119$ lines. This can be explained by analysing right-hand-side (RHS)
of Eq.(5). $E_{\parallel}=0$  at $t=0$ everywhere, however it can only be generated in the regions where $\partial_x E_x \not = 0$. The latter is true
only in the density gradient regions where $E_x$ becomes progressively oblique propagating.
Thus, Eq.(5), derived here for the first time, correctly explains the simulated process of $E_{\parallel}$ generation by IC waves.
 Also, note that $E_{\parallel}$ amplitude at time $t=16000 \omega_{ce}^{-1}$ attains value of 
$3\times10^{-5}$.
This is somewhat smaller value than the one obtained in the full kinetic (PIC) simulation \cite{tss05a,tss05b}. 
This is due to the different mass ratios: in the the kinetic (PIC) simulation $m_i/m_e=16$, but here it is 262.
In dimensional units this $E_{\parallel}$
corresponds to about 0.003 statvolt cm$^{-1}$ or 90 V m$^{-1}$, i.e. in such electric field electrons would be accelerated to the
energy of $\approx 10$ keV over the distance of 100 m. Note, however, that the generated $E_{\parallel}$ is 
oscillatory in space and time. The typical values of the Dreicer electric field on the corona is few $\times 10^{-3}$ 
V m$^{-1}$ \cite{t06b}, which implies the obtained $E_{\parallel}$ in our model exceeds the Dreicer value by  at least
four orders of magnitude, perhaps enabling the electron run away regime. This would imply that our model  is
more relevant to the acceleration of solar wind, rather than solving coronal heating problem.
Essentially acceleration of electrons would dominate over the heating as such. However, this seems uncertain because 
electron and ion fluid separation cannot go on forever, and some sort of discharge should eventually take place.
At any rate, similar kinetic simulations have shown \cite{glm04} (see their Fig.(11)) that 
that wave energy is converted into particle energy on timescales of $10^3 \omega_{pe}^{-1}$
(mind that the latter number is likely to be dependent on the mass ratio $m_i/m_e$).

In  Fig.(2) we present $(V_{ez}-V_{iz})$ which is proportional to $j_z$, the parallel (electron and ion {\it flow separation}) current.
It can be seen from this figure that $(V_{ez}-V_{iz})$ attains moderate values of $\approx 0.03c$.
\cite{glm04} stated the importance of {\it charge separation} before. However 
the cause of $E_{\parallel}$ generation is actually electron and ion flow separation (see below). The latter is quite different from
the electrostatic effect of charge separation, which is inherently a plasma kinetic effect. Electron and ion flow separation
is a fluid-like (non-kinetic) effect, because their distribution functions remain Maxwellian at all times.

In Fig.(3) we present of the final attained parallel electric field amplitude (within 3 periods of the driving ion cyclotron wave)
as a function of mass ratio, i.e. $E^*=\max(|E_z(x,z,t_{final})|)$ vs. $m_i/m_e$. 
We see that 
the amplitude attained by $E_{\parallel}$ decreases linearly with inverse of the mass ratio $m_i/m_e$.
The $x$-range in Fig.(3) is $m_i/m_e=30 - 1836$, so that rightmost point of the dashed line enables us to
grasp $E_{\parallel}$ for the case of realistic mass ratio (i.e. 1836). We thus gather that 
$E_{\parallel}=0.0085/1836= 4.630 \times 10^{-6}$ which  is $4.630 \times 10^{-6} \times E_0 = 4.7\times 10^{-4}$ statvolt cm$^{-1}$  or 
14 Vm$^{-1}$. Note that it is likely that the actual value will be even smaller if the dissipative
effects (resistivity, viscosity) will be taken into account. However, these are know to be very small in the 
solar coronal plasma. Thus, we do not expect a big change in this result.

In Fig. (4) we show what trends in generation of parallel (magnetic field aligned) electron and ion flow as a function of
the mass ratio. 
It follows from Fig.(4) that an increase in mass ratio does not have any effect on final 
parallel speed attained by electrons. However, the parallel ion
velocity decreases linearly with inverse of the mass ratio $m_i/m_e$. 
The ratio of final attained ion and electron flow amplitudes (within 3 periods of the driving ion cyclotron wave)
as a function of mass ratio, i.e. 
$V_* = \max(|V_{iz}(x,z,t_{final})|)/ \max(|V_{ez}(x,z,t_{final})|)$, shows clear scaling of $1/(m_i/m_e)$.

The conclusions that follow from the collective analysis of Figs.(3)--(4) initially may seem counterintuitive.
On one hand maximal attained $E_{\parallel}$ amplitudes drop off as $1/m_i$ (Fig.(3)). On the other hand, electron flow
maximal attained amplitudes do not depend on $m_i$ (they all are circa $0.03c$, see Fig.(4)), while 
ion flow maximal attained amplitudes are much smaller     ($0.0001c - 0.00065c$) than
that of electrons and drop off as $1/m_i$ (Fig.~(4)). Thus one might expect that more massive ions
should produce a bigger $E_{\parallel}$ (since separation of electron and ion fluids is the source
of $E_{\parallel}$ and {\it that} separation is expected to be largest in the case of more massive ions, as they are slower). 
In fact, this is what would be expected if the polarisation drift produced by the driving
IC wave is the cause of parallel electric field generation \cite{glq99,glm04}. 
The latter two references use the following polarisation drift current, $j_\perp=(m_i n_i /B^2) \partial E_\perp / \partial t$.
This equation implies that  $E_{\parallel}$ then should increase with $\propto m_i$.
However, in Fig.(3) we see completely opposite $E_{\parallel} \propto 1/ m_i$ scaling.
These results can be interpreted (reconciled) as following: (i) ion dynamics plays no role in the $E_{\parallel}$
generation, i.e. polarisation drift has no effect in contrary to the claims of \cite{glq99,glm04}; 
(ii) decrease in the generated parallel electric field amplitude with the increase of the mass ratio $m_i/m_e$
is caused by the fact that $\omega_d = 0.3 \omega_{ci} \propto 1/m_i$ is decreasing, and hence the electron fluid can
effectively "short-circuit" (recombine with) the slowly 
oscillating ions, hence producing smaller $E_{\parallel}$ which also scales exactly as $1/m_i$.

It should be noted that since here we use two-fluid approach the generated $E_{\parallel}$ cannot
change the distribution function, which obviously remains Maxwellian, while in the previous kinetic simulation of
a similar system it produced bumps in the distribution function as the electrons residing on the magnetic field lines with the density gradients
get efficiently accelerated (see e.g. Fig.(4) in \cite{tss05b}).

\begin{figure}
\centering
\resizebox{6.5cm}{!}{\includegraphics{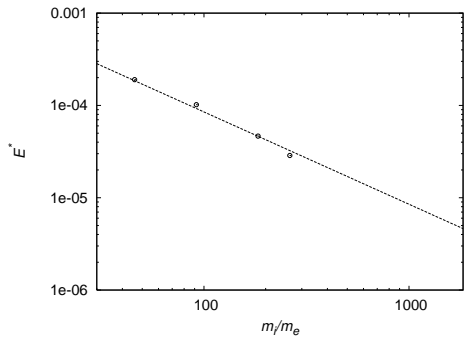} }
\caption[]{A log-log plot of the final attained 
parallel electric field amplitude (generated within 3 periods of the driving ion cyclotron wave)
as a function of mass ratio. Data points correspond to $E^*=\max(|E_z(x,z,t_{final})|)$ for
$m_i/m_e=$ 45.9, 91.8, 183.6 and 262.286.
The dashed line corresponds to the fit $ 0.0085 / (m_i/m_e)$.}
\end{figure}

\begin{figure}
\centering
\resizebox{6.5cm}{!}{\includegraphics{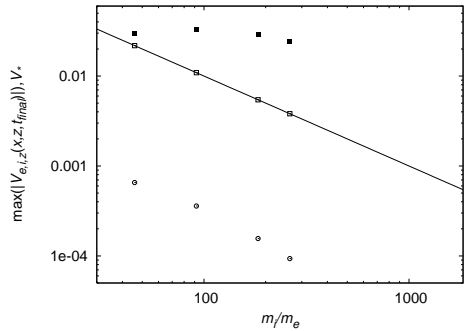} }
\caption[]{A log-log plot of the final attained 
electron and ion flow amplitudes (generated within 3 periods of the driving ion cyclotron wave)
as a function of mass ratio.
 Solid squares correspond to $\max(|V_{ez}(x,z,t_{final})|)$,
open circles to $\max(|V_{iz}(x,z,t_{final})|)$, while open squares show their ratio, i.e. 
$V_* = \max(|V_{iz}(x,z,t_{final})|)/ \max(|V_{ez}(x,z,t_{final})|)$. The solid line show a fit
$1 / x =1/(m_i/m_e)$.}
\end{figure}

The generation of $E_{\parallel}$  is a generic feature of plasmas with the transverse 
density inhomogeneity and in a different context this was known for decades in the
laboratory plasmas \cite{cm93,r82}. 
 Also, it should be emphasised that the two fluid description in the context of parallel electric field generation
has been used before \cite{gb79}. Moreover, \cite{l90} has explored similar equation. It can be easily shown
that our Eq.(5) can be reduced to Eq.(44) from \cite{l90}, by neglecting the displacement current and writing it in
Fourier-transformed form. At first sight, this may seem to diminish the importance of our result. However,
first, the inclusion of the displacement current mathematically means introduction of the time-dynamics ($\partial^2_{tt}$ term in Eq.(5))
which is crucial for the correct description of $E_{\parallel}$ generation by IC wave driving in the transversely inhomogeneous plasma;
second, generally the importance of the inclusion of the displacement current in this context has been recently stressed \cite{sl06};
and third, because of transverse inhomogeneity of plasma it is impossible to use Fourier-transform in the transverse spatial coordinate.
It should be noted when plasma density is homogeneous no $E_{\parallel}$ generation takes place,  in our model; and this is corroborated both by
 numerical simulations (not presented here) and agrees with the Eq.(5) (when $n=const$, the  RHS of Eq.(5) is zero at all times as $E_x$ does not 
 propagate obliquely).
 However, the relation of this claim to the observations is not entirely straightforward. 
Observations in the Earth's magnetosphere have for
many years noted that perpendicular gradients of the perpendicular
electric field are commonly seen during auroral zone crossings \cite{m77}. 
These are sometimes seen in the presence of perpendicular (transverse) density
gradients; and the
plasma density typically decreases above parallel potential drops \cite{c98}.
The question here is what is generated first $E_{\parallel}$ or $E_{\perp}$ (since obviously both are coupled via Eq.(5))?
Our model also correctly reproduces the previous kinetic results 
\cite{glq99,glm04,tss05a,tss05b} that only electrons are accelerated (along the background magnetic field),
while ions practically show no acceleration.  This applies only on time scales considered i.e. $t \leq 16000 / \omega_{ce}$. 
However, in the Earth's auroral zone a significant amount of ion acceleration takes place \cite{h82} and
Alfv\'enic Poynting flux is directly correlated with ion outflow in the auroral zone \cite{s00}. Based on this one can
conjecture that either processes reported by these observations acted over time scale $t\gg 16000 / \omega_{ce}$ so that ions due to their
large inertia have enough time to accelerate or a mechanism other then proposed here is responsible for the ion acceleration.
It should be noted that because of oscillatory nature of the obtained $E_{\parallel}$, it can possibly act as 
yet another mechanism for interpreting the peculiar hard x-ray ($> 25$ keV) solar flare, which is believed to be
produced by a non-thermal electron beam \cite{os06,vmn06}.

\begin{acknowledgments} Author is supported by  the University of Salford Research Investment Fund 2005 grant  
and Science and Technology Facilities Council  (UK) standard grant.
\end{acknowledgments}


\end{document}